\documentclass{aa}

\usepackage{graphicx}
\begin{document}
\title{Infrared spectroscopy of the largest known trans-neptunian object 
2001 KX$_{76}$}

\author{J. Licandro
 \inst{1}
 \and
 F. Ghinassi
	  \inst{1}
	  \and
	  L. Testi
	  \inst{2}
%	  \fnmsep
 }

\offprints{J. Licandro, e-mail licandro@tng.iac.es}

\institute{Centro Galileo Galilei \& Telescopio Nazionale Galileo, P.O.Box 565, E-38700, S/C de La Palma, Tenerife,
 		Spain.\\
	\and
	  INAF - Osservatorio Astrofisico di Arcetri, Largo E. Fermi 5, I-50125 Firenze,
	  Italy.\\
 }

  \date{Received February, 2002; accepted , 2002}

\abstract{
We report complete near-infrared (0.9-2.4 $\mu$m) spectral observations of 
the largest know trans-neptunian objects (TNO) 28976 = 2001 KX$_{76}$ 
taken in two different nights using the new Near Infrared Camera Spectrometer 
(NICS) attached to the 3.56m Telescopio Nazionale Galileo (TNG). 
The spectra are featureless and correspond to a neutral colored object.
Our observations indicate that the surface of 2001 KX$_{76}$ is probably
highly evolved due to long term irradiation, and that collisional resurfacing
processes have not played an important role in its evolution.
}

\titlerunning{Infrared spectroscopy of TNO 2001 KX$_{76}$}

\maketitle
%
%________________________________________________________________

\section{Introduction}

The trans-neptunian object (TNO) 28976 = 2001 KX$_{76}$ was discovered in May 2001,
and it is the largest known TNO (Millis et al. \cite{Millis}). Assuming a
geometric albedo of 4\% its diameter is $\sim$1100 km. TNOs are remnant 
planetesimals from the early solar system formation stages, populating a region 
just beyond the orbit of Neptune (Edgeworth \cite{Edgeworth49}; Kuiper 
\cite{Kuiper51}), called  the Edgeworth-Kuiper belt (EKb). They comprises three 
dynamical classes: objects in the 3:2 mean motion resonance with Neptune have 
been described as ``Plutinos'' (Jewitt \& Luu \cite{JewLuu96}), those beyond 
about $41\; AU$ as ``Classical EKb Objects'',
and those with a much larger semi-major axis and higher eccentricity than the 
previous classes are known as ``Scattered Disk Objects'' (Luu et al. 
\cite{Luu97}).
2001 KX$_{76}$ has orbital elements that fit with those of the Classical EKb Objects. 

The low temperature in the EKb and the low density of TNOs imply they are 
probably the most pristine objects in the Solar System. So they can provide key 
information on the composition and early conditions of the pre-solar nebula.
Though the study of their surface properties is very important from a 
cosmogonical point of view, as it could provide important clues to understand 
the conditions existing at the beginning of the solar system.

A powerful method for remote determination of the composition of volatile surface 
component of the outer solar system objects is the Near-infrared spectroscopy
(Brown \& Cruikshank \cite{BrownCrui97}).
Due to the small size and large distance of this objects, this technique is 
actually limited to the largest members of the EKb. Very few near-infrared 
spectra of TNOs have been already published: Brown et 
al. (\cite{Brownetal97}) present the spectrum of 1993 SC; Luu \& Jewitt 
(\cite{LuuJew98}) the spectrum of 1996 TL$_{66}$; Brown et al. (\cite{BrownCrPe99}) 
the spectrum of 1996 TO$_{66}$; Brown et al. (\cite{Brownetal2000}) the 
spectrum of 2000 EB$_{173}$; Licandro et al. (\cite{Lic}) the spectra of (20000)
Varuna (= 2000 WR$_{106}$) and 2000 EB$_{173}$; 
and Jewitt \& Luu (\cite{JewLuu2001}) the spectra of 2000
EB$_{173}$, 1999 DE$_{9}$, 1996 TS$_{66}$, and 1993 SC. 
Even the number of published spectra of TNOs is very low, large different 
surface properties have been inferred among them. Strong color variations, from
neutral to very red objects, have been also reported photometrically by several
authors (e.g. Jewitt \& Luu \cite{JewLuu2001}, Gil-Hutton \& Licandro
\cite{GilLic}, Davies et al. \cite{Davies00}, Barucci et al. 
\cite{Baruccietal00}). This color diversity is
confirmed in the reported infrared spectra. But also some absorption bands
possibly due to hydrocarbons and/or water ice were observed in the infrared
spectrum of some objects (1993 SC, 1999 DE$_{9}$, 1996 TO$_{66}$, 20000 Varuna), while
the spectra of other TNOs (1996 TL$_{66}$, 1996 TS$_{66}$, 2000 EB$_{173}$) are featureless 
in the observed range. This diversity of surface compositions has been observed
also in similar objects like Centaurs and irregular satellites 
(Brown \cite{Brown2000}). The observed surface diversity can be atributed to
intrinsically different compositions among TNOs and/or to some
resurfacing processes (Jewitt \& Luu \cite{JewLuu2001}; Gil-Hutton \cite{Gil}).

%__________________________________________________________________

\section{Observations}

We have obtained low resolution spectra of 2001 KX$_{76}$ on July 6.0 UT and
July 12.0 UT, 2001, with the 3.6m Telescopio Nazionale Galileo (TNG), 
using NICS, the near-infrared camera and spectrometer (see Baffa et al. 
\cite{Baf}). Among the many imaging and spectroscopic observing modes, NICS 
offers a unique, high throughput, low resolution spectroscopic mode with an 
Amici prism disperser (Oliva \cite{oliva01}), which yields a complete 0.9-2.4 
$\mu$m spectrum. A $1.5"$ width slit corresponding to a spectral resolving 
power $R\simeq34$ and quasi-constant along the spectrum, has been used.
The  low resolution together with the high efficiency of the Amici prism (about 
90\% across the useful wavelength range) allowed us to obtain spectra of faint objects 
like TNOs with a four meter class telescope for the first time (Licandro et al. 
\cite{Lic}), and with the advantage of having the whole infrared range measured 
simultaneously. 

The identification of the TNO was done by taking series of images through the 
J$_s$ filter separated by one our, and by comparing them. The object was 
identified as a moving object at the predicted position and with the predicted 
proper motion. The slit was oriented in the paralactic angle, and the tracking 
was at the TNO proper motion. The acquisition consisted of a series of 5 images 
of 60 seconds exposure time in one position (position {\em A}) of  the slit and 
then offsetting the telescope by $30"$ in the direction of the slit (position 
{\em B}). This process was repeated and a number of {\em ABBA} cycles were 
acquired. The total exposure time was 3600 seconds each night.
The reduction and calibration of the spectra was done as in Licandro et al.  
(\cite{Lic}).

\begin{figure*}
\centering
\includegraphics[angle=-90,width=\textwidth]{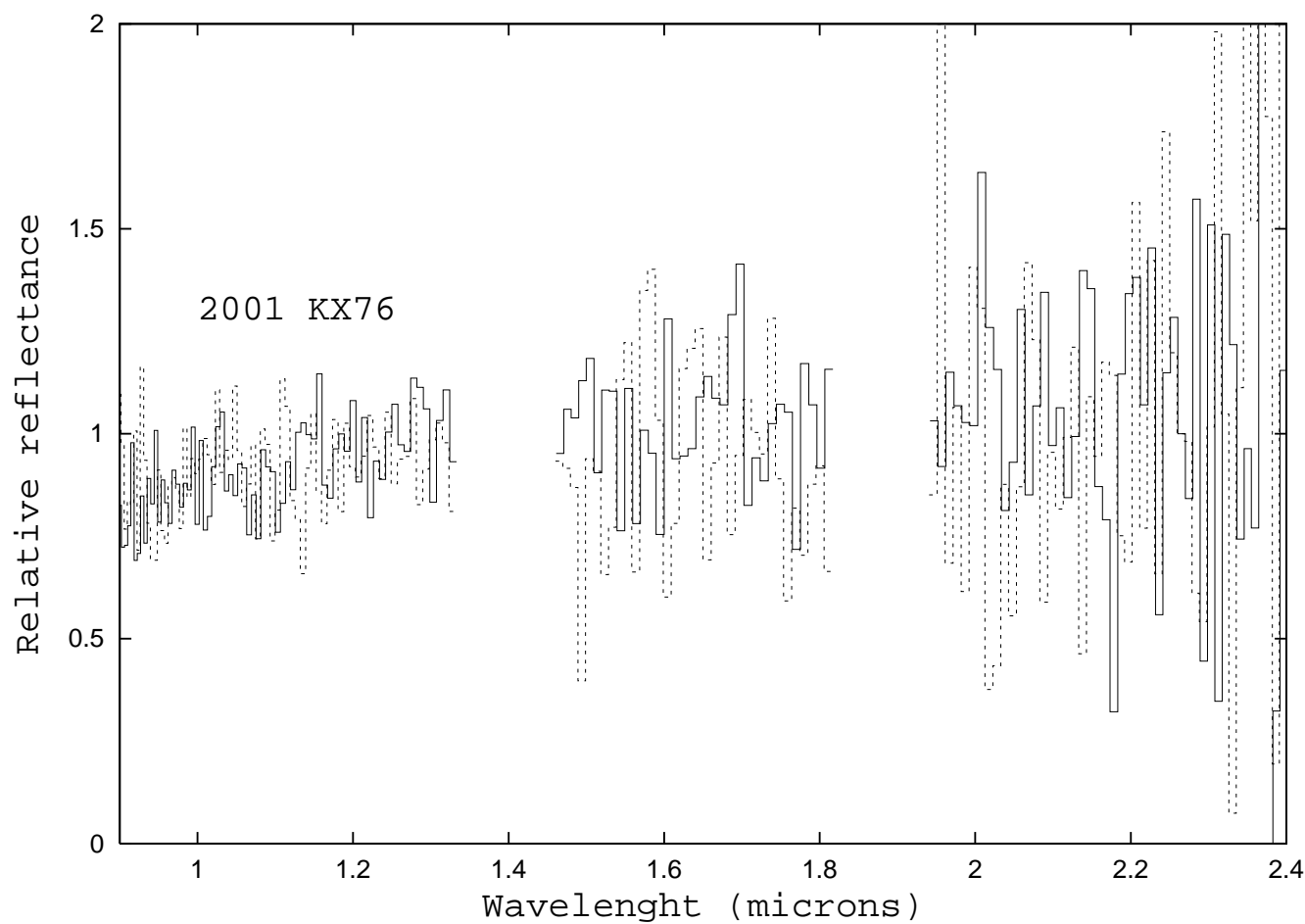}
\caption{Reflectance spectra of 2000 KX$_{76}$ obtained in two different nights.
Full line correspond to the spectrum obtained on July 6.0 UT, and
dashed line to the one taken on July 12.0 UT, 2001. The 
spectra has been normalized around 1.6 $\mu$m. Note that both spectra are 
identical withing the noise, featureles, and correspond to an object with 
neutral color. 
}
  \label{Fig1}
 \end{figure*}

To correct for telluric absorption and to obtain the relative reflectance, 
the G star Land (SA) 110-361 (Landolt \cite{Landolt}), which has visible colors 
very similar to that of the Sun, was observed during the same night just after 
2001 KX$_{76}$, and at a similar airmass. Land (SA) 110-361 was observed also in 
previous nights together with the solar analogue star P330E (Colina \& Bohlin 
\cite{Col}). Both stars present similar spectra in the 
infrared region, so we used Land (SA) 110-361 as a solar analogue. 
The spectrum of the TNO was divided by the spectrum of the solar analogue star, 
and then normalized to unity around 1.6 $\mu$m, thus obtaining the relative 
reflectance plotted in Fig. \ref{Fig1}. Around the telluric water band 
absorptions the S/N of the spectrum is very low, additionally, the telluric absorption 
varies between the TNO spectra and the standard stars spectra introducing false 
spectral features. Therefore, these parts are not included in the final spectra.

%\input{table1}

% Two column figure (place early!)
%______________________________________________ Gamma_1 (lg rho, lg e)
%
% Two column figure (place early!)
%______________________________________________ Gamma_1 (lg rho, lg e)
\begin{figure*}
\centering
\includegraphics[angle=-90,width=\textwidth]{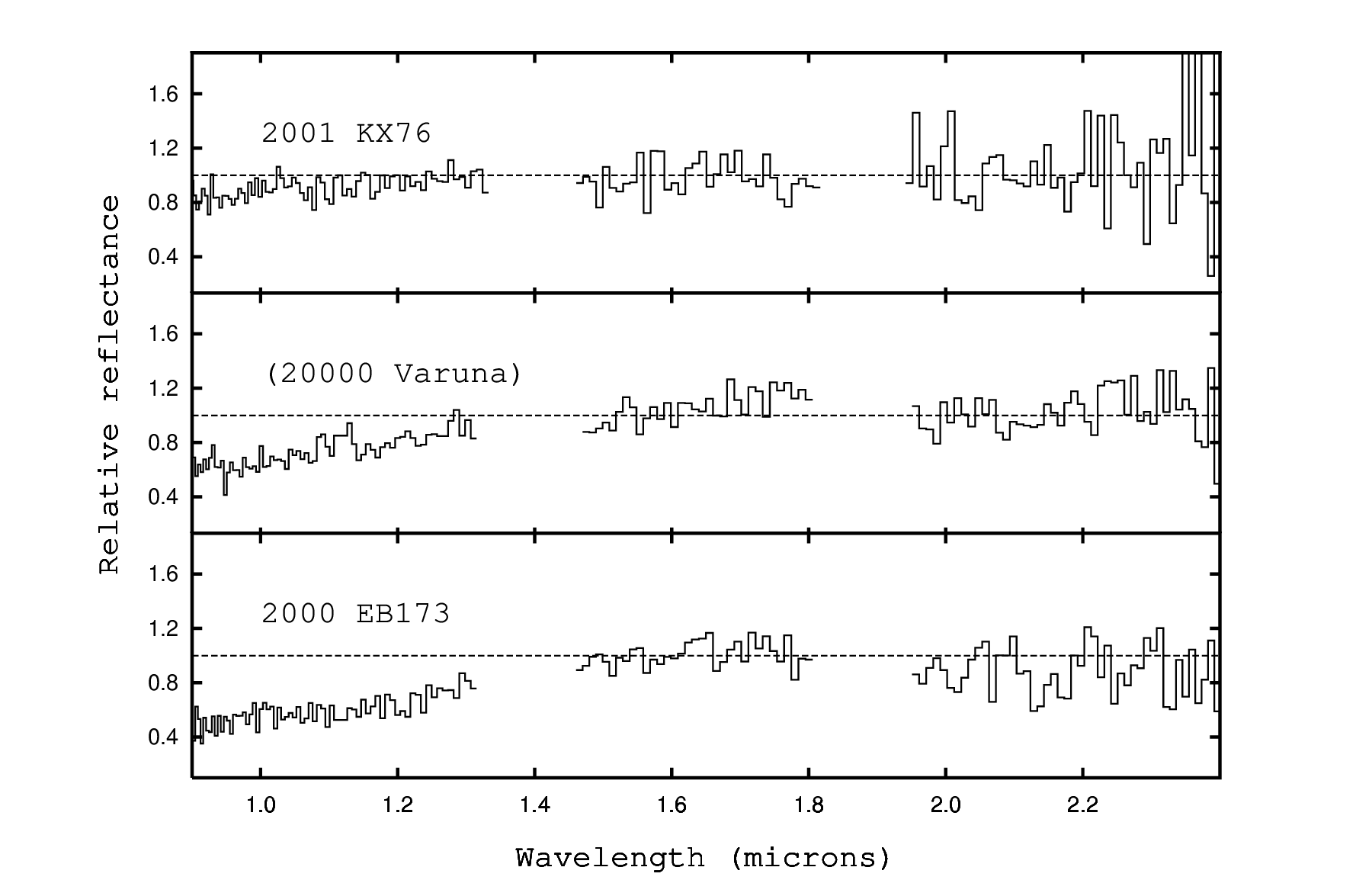}
\caption{The reflectance spectrum of 2001 KX$_{76}$ is ploted together with 
the spectra of  Varuna and 2000 EB$_{173}$ taken from Licandro et al.
2001. The spectrum of 2001 KX$_{76}$ is the mean of the spectra presented in 
Fig. 1. 
Note that the spectrum of 2001 KX$_{76}$ do not present the strong water ice 
absorption bands observed in the spectrum of  Varuna.
}
  \label{Fig2}
 \end{figure*}
%
 
%__________________________________________________________________

\section{Discussion}

The spectra of 2001 KX$_{76}$ are shown in Fig. \ref{Fig1}. 
Both are identical withing the noise. 
This means that the object was observed at a very similar rotational phase or 
that the composition of its surface is very homogeneous.
The spectrum of 2001 KX$_{76}$ is featureless and,
compared with the very red slope spectra of Varuna and 
2000 EB$_{173}$ (Licandro et al. \cite{Lic}), is almost neutral (see Fig. 
\ref{Fig2}).

The spectrum of 2001 KX$_{76}$ do not show the water ice absorption features at 
1.5 and 2.0 $\mu$min observed in Varuna. 
Similar objects with neutral colors present a large variation in the 
detected amount of water ice in the surface. Brown et al. (\cite{BrownCrPe99}) 
detected water ice in 1996 TO$_{66}$ while Luu \& Jewitt (\cite{LuuJew98}) did 
not in 1996 TL$_{66}$. Also red TNOs presents  this variation in the
detected amount of water ice in their surfaces (see Licandro et al. \cite{Lic}).
Though it seems that the diversity between red and 
neutral TNOs has no relation with the presence of water ice in the surface. 

The observed very red color of some TNOs is probably due to an evolved surface 
which is the result of long term irradiation by solar radiation, solar wind, 
and galactic cosmic-rays. This results in the selective loss of hydrogen and the
formation of an ``irradiation mantle" of carbon residues (Moore et al. 
\cite{Mooreetal83}; Johnson et al. \cite{Johnetal84}; Strazzulla \& Johnson 
\cite{StrazzullaJo91}). This process makes that an initially neutral color and 
high albedo ice becomes redish. But further irradiation gradually reduces the 
albedo at all wavelengths, and the material becomes very dark, neutral in color,
and spectrally featureless (Andronico et al. \cite{Androetal87}; Thompson et al.
\cite{Thompsonetal87}). Gil-Hutton (\cite{Gil}) shows that the total amount of
radiation received by TNOs during the age of the solar system can produce
very dark, neutral surfaces. The neutral color and the lack of ice features in
the spectrum of 2001 KX$_{76}$ suggest that the surface of this object has a
very evolved irradiation mantle, that has not suffered from a strong collisional
resurfacing process. In contrary, the presence of water ice in the surface of
Varuna suggest that it has undergone a strong resurfacing due to collisions,
capable of retrieve ``fresh" material from below the irradiation mantle. 
Even if the collisional resurfacing process were not the
primary cause of the observed color dispersion of the TNOs according to
Jewitt \& Luu (\cite{JewLuu2001}), it could be the reason to explain why
water ice is observed in the very evolved surface of these objects.
Thus, more infrared spectroscopy of a large number of TNOs is needed.

\section{Conclusions}

Two spectra of 2001 KX$_{76}$, obtained in different nights, are presented.
Both are almost identical withing the noise, suggesting that the TNO 
was observed at a very similar rotational phase or 
that the composition of its surface is very homogeneous.
The spectra are featureless and correspond to a neutral colored object. This
suggest that the surface of 2001 KX$_{76}$ is probably highly evolved by long
term radiation and that any possible collisional resurfacing process did not
erode significantly its irradiation mantle.

\begin{acknowledgements}
JL thanks Bobby Bus and Humberto Campins for their usefull suggestions on faint
solar analogue stars to be observed for calibrations purposes with large 
telescopes.
 
This paper is based on observations made
with the Italian Telescopio Nazionale Galileo (TNG) 
operated on the island of La Palma by the Centro Galileo
Galilei of the CNAA (Consorzio Nazionale per l'Astronomia e l'Astrofisica) 
at the Spanish Observatorio del
Roque de los Muchachos of the Instituto de Astrofisica de Canarias.
We are grateful to all the technical staff and telescope operators for
their assistance during the commissioning phase of NICS.
\end{acknowledgements}

%%%%%%%%%% Bibliografia %%%%%%%%%%%%%%%%%%%%%

\end{document}